# Localized Surface Plasmon Resonance on Two-Dimensional HfSe$_2$ and ZrSe$_2$

Hemendra Nath Jaiswal,[1] Maomao Liu,[1] Simran Shahi,[1] Fei Yao,[2] Qiyi Zhao,[3] Xinlong Xu,[3, a)] and Huamin Li[1, a)]

[1] *Department of Electrical Engineering, University at Buffalo, the State University of New York, Buffalo, New York 14260, US*

[2] *Department of Materials Design and Innovation, University at Buffalo, the State University of New York, Buffalo, New York 14260, US*

[3] *Shaanxi Joint Lab of Graphene, State Key Lab Incubation Base of Photoelectric Technology and Functional Materials, International Collaborative Center on Photoelectric Technology and Nano Functional Materials, Institute of Photonics & Photon-Technology, Northwest University, Xi'an 710069, P.R. China*

[a)] Authors to whom correspondence should be addressed. Electronic address: xlxuphy@nwu.edu.cn and huaminli@buffalo.edu

**Abstract**

HfSe$_2$ and ZrSe$_2$ are newly discovered two-dimensional (2D) semiconducting transition metal dichalcogenides (TMDs) with promising properties for future nanoelectronics and optoelectronics. We theoretically revealed the electronic and optical properties of these two emerging 2D semiconductors, and evaluated their performance for the application of localized surface plasmon resonance (LSPR) at extreme conditions – in-plane direction versus out-of-plane direction and monolayer versus multilayer. First, the energy band structure and dielectric constants were calculated for both the monolayer and multilayer structures using Kohn-Sham density functional theory (KS-DFT) with van der Waals (vdW) corrections. A parallel-band effect observed in the monolayer band structure indicates a strong light-matter interaction. Then, based on the calculated dielectric constants, the performance of the LSPR excited by Au sphere nanoparticles (NPs) was quantitatively characterized, including polarizability, scattering and absorption cross-sections, and radiative efficiency using Mie theory. For the multilayer HfSe$_2$ and ZrSe$_2$, the LSPR showed very comparable intensities in both the in-plane and out-of-plane directions, suggesting an isotropy-like light-matter interaction. In a comparison, the LSPR excited on the monolayer HfSe$_2$ and ZrSe$_2$ was clearly observed in the in-plane direction but effectively suppressed in the out-of-plane direction due to the unique anisotropic nature. In addition to this extraordinary anisotropy-to-isotropy transition as the layer number increases, a red-shift of the LSPR wavelength was also found. Our work has predicated the thickness-dependent anisotropic light-matter interaction on the emerging 2D semiconducting HfSe$_2$ and ZrSe$_2$, which holds great potential for broad optoelectronic applications such as sensing and energy conversion.

**I. Introduction**

Localized surface plasmon resonance (LSPR) has been considered as one of the most promising approaches for achieving tunable light concentration, such as nanoscale sensing [1], light-to-electricity and light-to-heat energy conversion/harvesting [2, 3]. The LSPR is frequently excited by electromagnetic (EM) radiation in proximate metallic nanostructures, which causes selective photon extinction and local EM field enhancement. The dependence of the LSPR on the size, shape, and material of the metallic nanostructures as well as on the dielectric environments has been studied extensively [1, 4-6]. For example, $SiO_2$ is the most commonly used dielectric material for realizing LSPR as compared to other dielectrics, because it is a good passivation layer, grown easily by thermal oxidation, and compatible with most semiconductor fabrication processes. Our previous work has investigated the LSPR excited on bulky high-dielectric-constant ($k$) materials including $Si_3N_4$, $ZrO_2$, and $HfO_2$, and has demonstrated the advantages of using these high-$k$ dielectric materials, such as increased absorption efficiency, enlarged cross-section area, and enhanced transmission of the EM field [7, 8].

With the rise of graphene [9, 10], two-dimensional (2D) materials have received extensive attention since 2004, due to their unique quantum confinement which possesses various exceptional properties compared to their bulk counterparts [11-14]. The 2D material family, including graphene, phosphorene, transition metal dichalcogenides (TMDs), and hexagonal boron nitride ($h$-BN) *etc.*, exhibit diverse electronic properties ranging from semimetal to semiconductor, and to insulator. This wide bandgap variation, from zero to about 6 eV, covers a broad spectrum from ultraviolet (UV) to far-infrared (FIR). Together with their strong light-matter interaction, unique layer number dependence, as well as numerous possibilities of van der Waals (vdW) hetero-stacking combination, the 2D materials are capable of realizing various novel photonic devices for energy conversion between light and other energy forms. More recently, group IVB-VIA 2D TMDs such as HfSe$_2$ and ZrSe$_2$ layered crystals have been found to be promising for future high-performance nanoelectronics and optoelectronics, due to their comparable band gaps (1.2 eV for the monolayer and 0.9 for the multilayer [15-19]) with Si, remarkably high phonon-limited mobility (~3500 cm$^2$/Vs for HfSe$_2$ and ~2300 cm$^2$/Vs for ZrSe$_2$ at room temperature [17]), and unique capability of forming native high-*k* oxides (HfO$_2$ and ZrO$_2$) [18, 19]. Given the fact that most of the studies on 2D HfSe$_2$ and ZrSe$_2$ only focus on the in-plane properties so far [19-23], future investigation and application of 2D HfSe$_2$ and ZrSe$_2$ should consider the nature of their anisotropy and explore in both the in-plane and out-of-plane directions.

In this work, we theoretically evaluated the LSPR excited by Au sphere nanoparticles (NPs) on both 2D multilayer and monolayer HfSe$_2$ and ZrSe$_2$, and quantitatively compared the in-plane and out-of-plane light-matter interaction properties. We first calculated the energy band structures and dielectric constants using Kohn-Sham density functional theory (KS-DFT) with the vdW corrections. A strong light-matter interaction was predicted due to the parallel-band effect observed in the monolayer band structure. Then, based on the calculated dielectric constants, we quantitatively evaluated the performance of the LSPR excited by Au sphere NPs, including polarizability, scattering and absorption cross-sections, and radiative efficiency using Mie theory. The LSPR excited on the multilayer HfSe$_2$ and ZrSe$_2$ showed very comparable intensities in both the in-plane and out-of-plane directions, suggesting an isotropy-like light-matter interaction. In contrast, the LSPR on the monolayer HfSe$_2$ and ZrSe$_2$ was clearly observed in the in-plane direction but effectively suppressed in the out-of-plane direction, indicating a strong anisotropic interaction. In addition to this unique anisotropy-to-isotropy transition as the layer number increases, a red-shift of the LSPR wavelength was also found. Our work has predicated the LSPR performance on the emerging 2D semiconducting HfSe$_2$ and ZrSe$_2$ layered crystals at extreme conditions (in-plane direction versus out-of-plane direction and monolayer versus multilayer), which can be explored further for broad optoelectronics applications.

**II. Computational methodology**

Both HfSe$_2$ and ZrSe$_2$ can be represented as MX$_2$, where M is transition metal (Hf or Zr) and X is chalcogen (Se). A crystal structure of MX$_2$ is shown in Fig. 1, where a monolayer MX$_2$ consists of one layer of M atoms sandwiched by two layers of X atoms with a strong ionic bond in a trigonal structure [20, 24]. The interlayer bonding between each MX$_2$ layer is dominated by the vdW interaction. Because of the 2D layered structure, MX$_2$ has a strong anisotropic nature: the in-plane properties governed by the lattice basis vectors ***a*** and ***b***, and the out-of-plane properties governed by the lattice basis vector ***c***. Accordingly, the electronic and photonic properties, such as the dielectric function (or refractive index), can be categorized into two parts: the in-plane one which is perpendicular to the vector ***c*** as $\varepsilon_\perp$, and the out-of-plane one which is parallel to the vector ***c*** as $\varepsilon_\parallel$ [24, 25].

The electronic properties of HfSe$_2$ and ZrSe$_2$ were obtained by KS-DFT with the vdW forces schemed by Grimme and Tkatchenko *et al*. [26, 27]. The detailed description can be found in our previous work [24]. In brief, the projector augmented wave (PAW) scheme as implemented in the Vienna *ab initio* Simulation Package (VASP) was employed to mimic electron-ion interaction [28, 29]. The Perdew-Burke-Ernzerhof (PBE) parameterization of the generalized gradient approximation (GGA) [30, 31] which was corrected by Green's function and screened Coulomb interaction approximation (GWA) was adapted as exchange correlation for the calculations. $9 \times 9 \times 9$ $k$-mesh based on gamma-centered scheme was set for sampling the Brillouin zones with an energy cut-off of 400 eV. The Hellmann-Feynman force between each atom was set to be less than 0.01 eV/Å, and the relaxation of energy was set as $10^{-5}$ eV. For the study of dielectric properties of materials, the Heyd-Scuseria-Ernzerhof (HSE06) calculations were performed within the framework of PAW method [32]. Following the determination of the electronic ground state by hybridization calculations, the energy-dependent dielectric matrix with the imaginary part of the dielectric constants in PAW methodology was obtained [33]. The real part of the dielectric constants was obtained by Kramers-Kronig relations [34]. The lattice parameters of the structure were obtained from experimental values and passed the converge test [24]. The monolayer was generated by constructing a space wide enough between adjacent layers in *z*-direction of the bulk structure. A vacuum region of 30 Å was used to isolate the layers along *c* axis and was sufficient to eliminate the interaction between the adjacent layers.

### III. Results and discussion

*Energy band structures and dielectric properties*

The band structures of multilayer $HfSe_2$ and $ZrSe_2$ obtained by the PBE method are shown in Fig. 2 (a) and (b). They both are indirect-band-gap semiconductors with conduction band minima (CBM) at M of the high symmetry *k*-points and valence band maxima (VBM) at Γ of the high symmetry *k*-points. The calculated band gap values of the multilayer $HfSe_2$ and $ZrSe_2$ equal to 1.07 and 1.08 eV, respectively, which are reasonably consistent with the reported theoretical and experimental data (1.08-1.15 eV for $HfSe_2$ [35-37] and 1.10-1.20 eV for $ZrSe_2$ [35, 38-40]). The slight differences are due to the different ground state structures or pseudopotentials used for the calculations. For the monolayer $HfSe_2$ and $ZrSe_2$, the energy band structures are shown in Fig. 2 (c) and (d). The energy band gap value is 0.68 eV for monolayer $HfSe_2$ and 0.42 eV for monolayer ZrSe$_2$. Compared to conventional TMD semiconductors such as MoS$_2$ and WSe$_2$, it is found that HfSe$_2$ and ZrSe$_2$ have their own unique properties. First, the monolayer HfSe$_2$ and ZrSe$_2$ are also the indirect-band-gap semiconductors which are identical to their bulky counterparties [35]. Whereas MoS$_2$ and WSe$_2$ have the direct band gaps in their monolayer structures. Second, the monolayer HfSe$_2$ has the energy band gap larger than that of monolayer ZrSe$_2$, while the band gap of multilayer ZrSe$_2$ is larger than that of multilayer HfSe$_2$. The difference in the band gaps between monolayer and multilayer structures may be attributed to the electronegativity of bonded atoms and the interaction between layers [41]. For the same anions, the value of band gap is inversely proportional to the electronegativity of cations. Besides, the layer-to-layer interaction also affects the states of bonded atoms and reduces the band gaps of the multilayer structures [42]. Third, the conduction and valence bands are found to be parallel with each other around Γ and M of the high symmetry $k$-points in the monolayer HfSe$_2$ and ZrSe$_2$. This parallel-band effect in the monolayer structures can induce the Van Hove singularity (saddle points of joint density of states) and thus the maxima in the absorption spectra [43-45] which suggests the strong light-matter interaction.

Assuming that the complex dielectric function is defined as $\varepsilon = \varepsilon'+i\varepsilon''$, the real parts of the dielectric constants ($\varepsilon'$) for the multilayer HfSe$_2$ and ZrSe$_2$ are obtained as a function of the wavelength ($\lambda$) using Kramers-Kronig relations [34], as shown in Fig. 3 (a) and (b). Because of the anisotropic nature, the real part of the dielectric function which is perpendicular to the lattice basis vector ***c*** is denoted as the in-plane $\varepsilon'$ ($\varepsilon'_\perp$) and the one parallel to the lattice basis vector ***c*** is denoted as the out-of-plane $\varepsilon'$ ($\varepsilon'_{//}$). For both the multilayer HfSe$_2$ and ZrSe$_2$, the maximum of $\varepsilon'_\perp$ is larger than 15 which is over 2 times larger than that of $\varepsilon'_{//}$ (approximately 7). In a comparison, the maximums of $\varepsilon'_\perp$ and $\varepsilon'_{//}$ are reduced to approximately 4 and 1, respectively, in both the monolayer

HfSe$_2$ and ZrSe$_2$, as shown in Fig. 3 (c) and (d). Especially, with the change of the wavelength, the variation of $\varepsilon'_\parallel$ is much smaller compared to that of $\varepsilon'_\perp$, indicating a strong anisotropy of the dielectric properties in the monolayer structure.

The imaginary parts ($\varepsilon''$) of the dielectric constants of HfSe$_2$ and ZrSe$_2$ were calculated by HSE06, as shown in Fig. 4. Both the multilayer HfSe$_2$ and ZrSe$_2$ show a similarity of $\varepsilon''$ between the in-plane ($\varepsilon''_\perp$) and out-of-plane ($\varepsilon''_\parallel$) values when the wavelength is larger than ~1,000 nm or, in another word, the energy is less the band gap value, as shown in Fig. 4 (a) and (b). A clear discrepancy can be observed for the shorter wavelength than 1,000 nm which corresponds to the energy range larger than the band gap value. These energy-dependent variations are attributed to the interband transition and are in a good agreement with the previous experiments [36, 38, 46] that indicated the indirect transition from the valence bands to the first conduction bands. The maximums for $\varepsilon''_\perp$ and $\varepsilon''_\parallel$ are found to be approximately 23 and 10, respectively, in the multilayer HfSe$_2$ and ZrSe$_2$. As a comparison, the peak values of the monolayer $\varepsilon'_\perp$ and $\varepsilon'_\parallel$ are decreased significantly, as shown in Fig. 4 (c) and (d). It is also noted that the peaks of the monolayer $\varepsilon''_\perp$ are sharper than that of the multilayer ones, suggesting a strong light-matter interaction in the monolayer structures. This is consistent with the parallel-band effect as observed in the energy band structures. Considering both the multilayer and monolayer cases, the peaks of the dielectric constants are mainly attributed to the transitions among the first, second, third valence bands and the first conduction bands. The direction of transitions is from Γ to M, M to K, and K to Γ of the high symmetry *k*-points [24].

*LSPR characterizations*

Conventionally the LSPR is excited by the metal (*e.g.*, Au or Ag *etc.*) NPs placed on semiconductor surface with an incident light illustrating on the top through the air. To simplify this geometric structure and make Mie scattering formalism applicable, here we establish a physical model where an Au NP is embedded in a 2D material medium, and assume two extreme conditions – the incident light aligns in the in-plane direction and in the out-of-plane direction, as shown in Fig. 5. The 2D material medium includes the multilayer HfSe$_2$ and ZrSe$_2$, as well as their monolayer structures. For the monolayer case, the established model is approximately analogous to the structure where an Au NP placed on the top of a monolayer 2D semiconductor with the incident light illustrating in the in-plane or out-of-plane direction. This type of structure has been broadly adopted in various 2D materials such as graphene [47, 48], MoS$_2$ [49, 50], and WS$_2$ [51] *etc.*

In Mie theory, the polarizability of a particle describes the charge distribution behavior under an external electric field. Assuming a spherical metallic NP has a radius of *a*, its polarizability (*α*) can be expressed as [52]:

$$\alpha = 4\pi a^3 \frac{\varepsilon_p(\lambda) - \varepsilon_m(\lambda)}{\varepsilon_p(\lambda) + 2\varepsilon_m(\lambda)} \qquad (1)$$

where $\varepsilon_p(\lambda) = \varepsilon_p'(\lambda) + i\varepsilon_p''(\lambda)$ and $\varepsilon_m(\lambda) = \varepsilon_m'(\lambda) + i\varepsilon_m''(\lambda)$ are the wavelength-dependent complex dielectric constants of the metal particle and the dielectric medium, respectively. The absolute value of *α* is calculated for both the multilayer and monolayer HfSe$_2$ and Zrse$_2$, as shown in Fig. 6. For an Au [53] NP with *a* = 75 nm, the multilayer HfSe$_2$ can induce a peak of |*α*| to about 4.01 × 10$^{-20}$ and 2.47 × 10$^{-20}$ m$^3$ in the in-plane and out-of-plane directions, respectively. In contrast, the monolayer HfSe$_2$ shows a peak of |*α*| to about 1.31 × 10$^{-20}$ m$^3$ in the in-plane direction but only about 7.02 × 10$^{-21}$ m$^3$ in the out-of-plane direction. In addition, a red-shift of the polarizability peak, from 616.9 to 756.1 nm for the in-plane one and from 521.0 to 659.6 nm for the out-of-plane one, is observed with the increasing layer number of HfSe$_2$ from the monolayer to the multilayer. For the case of ZrSe$_2$, the similar results are also obtained.

The metallic NPs scatter and absorb the light to its maximum capacity at the LSPR wavelength because of combined oscillations of the electrons [6]. When a single metallic NP has a very small size with respect to the incident wavelength, the scattering cross-section ($C_{sca}$) and absorption cross-section ($C_{abs}$) of a dipole model are defined by the polarizability as [52]:

$$C_{sca} = \frac{1}{6\pi}\left(\frac{2\pi}{\lambda}\right)^4 |\alpha|^2 = \frac{8\pi a^6}{3}\left(\frac{2\pi}{\lambda}\right)^4 \left|\frac{\varepsilon_p(\lambda) - \varepsilon_m(\lambda)}{\varepsilon_p(\lambda) + 2\varepsilon_m(\lambda)}\right|^2 \qquad (2)$$

$$C_{abs} = \frac{2\pi}{\lambda}\text{Im}[\alpha] = 4\pi a^3 \left(\frac{2\pi}{\lambda}\right)\text{Im}\left[\frac{\varepsilon_p(\lambda) - \varepsilon_m(\lambda)}{\varepsilon_p(\lambda) + 2\varepsilon_m(\lambda)}\right] \qquad (3)$$

The sum of both the scattering and absorption is known as light extinction, and its corresponding cross-section is expressed as $C_{ext} = C_{sca} + C_{abs}$. Both $C_{sca}$ and $C_{abs}$ as functions of $a$ are plotted at the LSPR wavelength of 620 nm which corresponds to the photon energy of 2 eV, as shown in Fig. 7. For the case of the multilayer HfSe$_2$, $C_{sca}$ is smaller than $C_{abs}$ when $a$ is less than 75 nm in both the in-plane and out-of-plane directions, indicating that the light extinction is dominated by the light absorption. Most of the photon energy is dissipated in heat which can be used for solar glazing and nanoscale lithography. When $a$ is larger than 75 nm, $C_{sca}$ is larger than $C_{abs}$, suggesting the dominance of the light scattering which can be used for light trapping and light concentration. For the multilayer ZrSe$_2$, this threshold radius that implies the boundary of the scattering dominance and absorption dominance is 105 nm. Moreover, the threshold radii for the multilayer structure are consistent in both the in-plane and out-of-plane directions, suggesting an isotropy-like light-matter interaction. As a comparison, for the monolayer HfSe$_2$ and ZrSe$_2$, the threshold radius in the out-of-plane direction is significantly reduced than that in the in-plane direction. The difference of the threshold radii indicates a clear anisotropic response in the monolayer structure.

This unique thickness-dependent anisotropic light-matter interaction can be further confirmed by the radiative efficiency or scattering efficiency which is defined as $C_{sca}/C_{ext}$, as shown in Fig. 8. First, the multilayer $HfSe_2$ has the very comparable radiative efficiencies in both the in-plane and out-of-plane directions, which are ~50% at $a = 75$ nm. In contrast, the monolayer $HfSe_2$ shows a much higher out-of-plane radiative efficiency in comparison to the in-plane value. Second, the isotropy-like light-matter interaction on the multilayer $HfSe_2$ and the anisotropic interaction on the monolayer $HfSe_2$ occur in both the absorption-dominated (small $a$) and scattering-dominated (large $a$) segments. Even at $a = 100$ nm, the out-of-plane radiative efficiency of the monolayer $HfSe_2$ is still higher than that in the in-plane direction. Similar to $HfSe_2$, the thickness-dependent anisotropic light-matter interaction on $ZrSe_2$ is also obtained.

Anisotropy of NP-induced plasmon is a field of particular interest. For example, Qiu *et al.* [54] systematically reviewed the plasmonic effect for anisotropic particles with rectangular and spherical anisotropic coordinates. Razumova *et al.* [55] discussed the solution for the plasmonic effect induced by anisotropic dielectric mediums. As a comparison, our strategy in this work is to evaluate and compare the LSPR at the extreme conditions – the in-plane direction versus the out-of-plane direction and the monolayer versus the multilayer. The model used in this work is simplified, yet it has limitations. For example, the estimation of the LSPR was exclusively based on the direction dependence and layer number dependence of the dielectric functions of 2D materials. Therefore, the results might be overestimated due to the assumption set at the extreme conditions. Experimental works or numerical simulations are needed to further explore the unique anisotropy of 2D materials and the related light-matter interaction.

**IV. Conclusion**

In summary, we have revealed the strong isotropy-like light-matter interaction with respect to the in-plane and out-of-plane directions on the multilayer $HfSe_2$ and $ZrSe_2$, as well as the strong anisotropic response on the monolayer $HfSe_2$ and $ZrSe_2$. We first calculated the anisotropic dielectric functions from KS-DFT calculation with the vdW corrections, then estimated the performance of the LSPR excited by the Au NP on $HfSe_2$ and $ZrSe_2$, in terms of the polarizability, scattering and absorption cross-section, and radiative efficiency using Mie theory.

**Acknowledgments**

This work was supported by the National Science Foundation (NSF) under Grant No. 1745621.

**Figures**

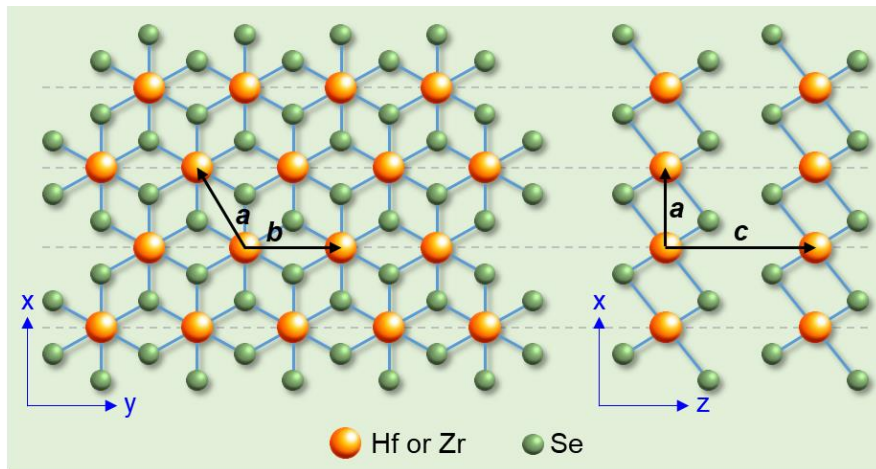

FIG. 1. Atomic crystal structure of 2D $MX_2$ ($HfSe_2$ or $ZrSe_2$) in top view (left) and side view (right).

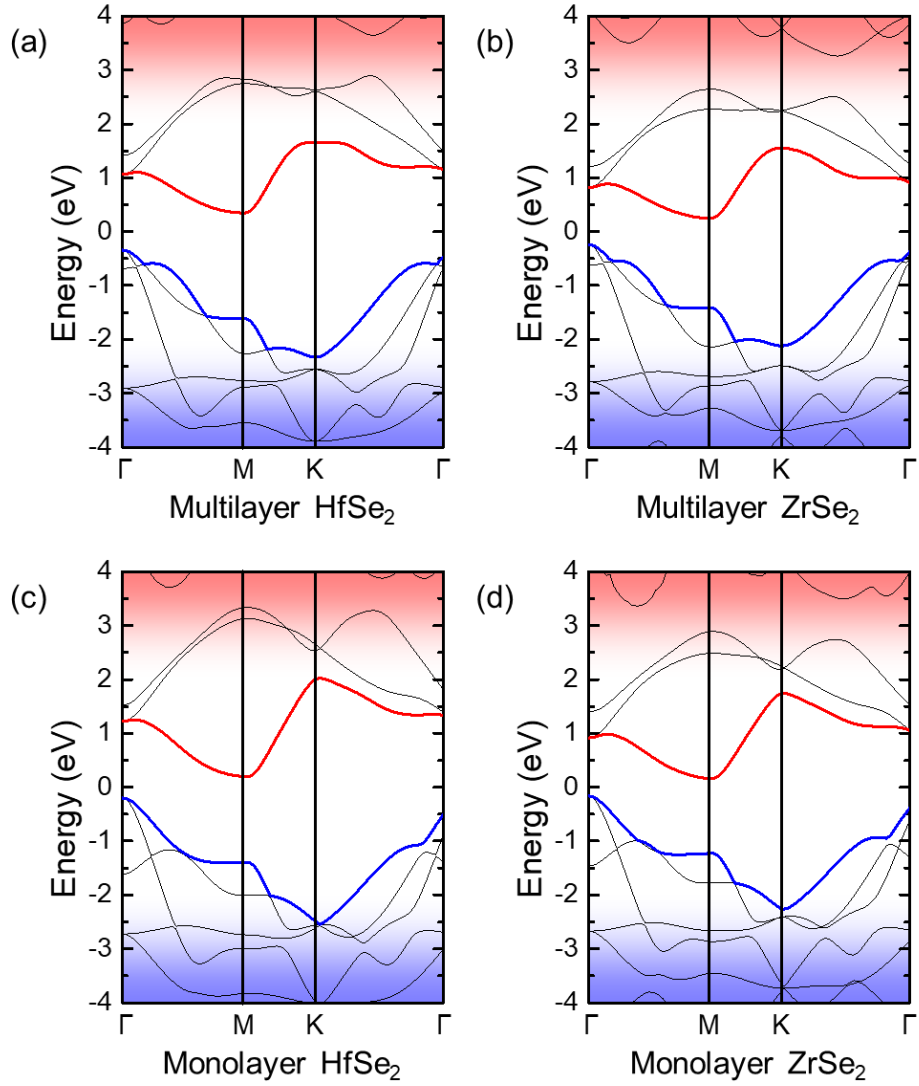

FIG. 2. Calculated energy band structures of (a) multilayer $HfSe_2$, (b) multilayer $ZrSe_2$, (c) monolayer $HfSe_2$, and (d) monolayer $ZrSe_2$. The red and blue lines denote the lowest conduction band edge and the highest valence band edge, respectively.

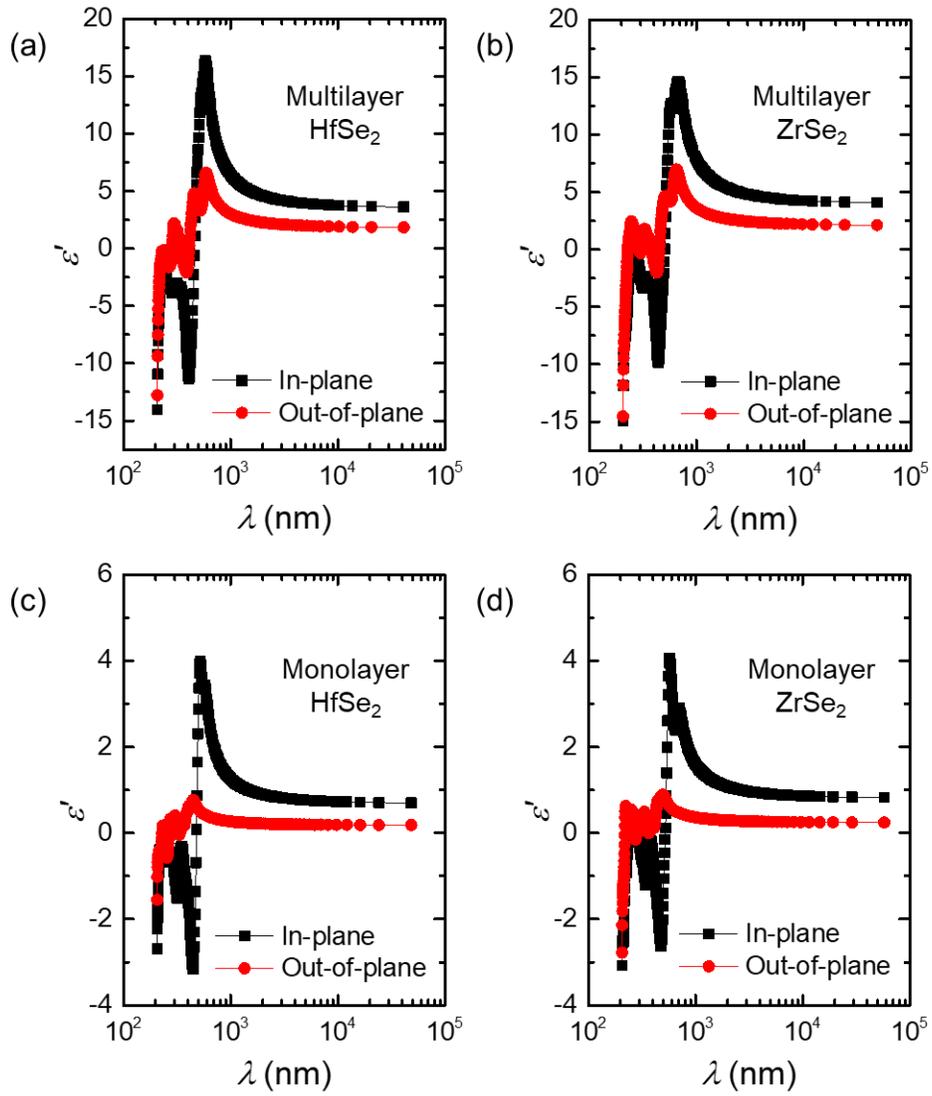

FIG. 3. Real parts of dielectric constants of (a) multilayer $HfSe_2$, (b) multilayer $ZrSe_2$, (c) monolayer $HfSe_2$, and (d) monolayer $ZrSe_2$.

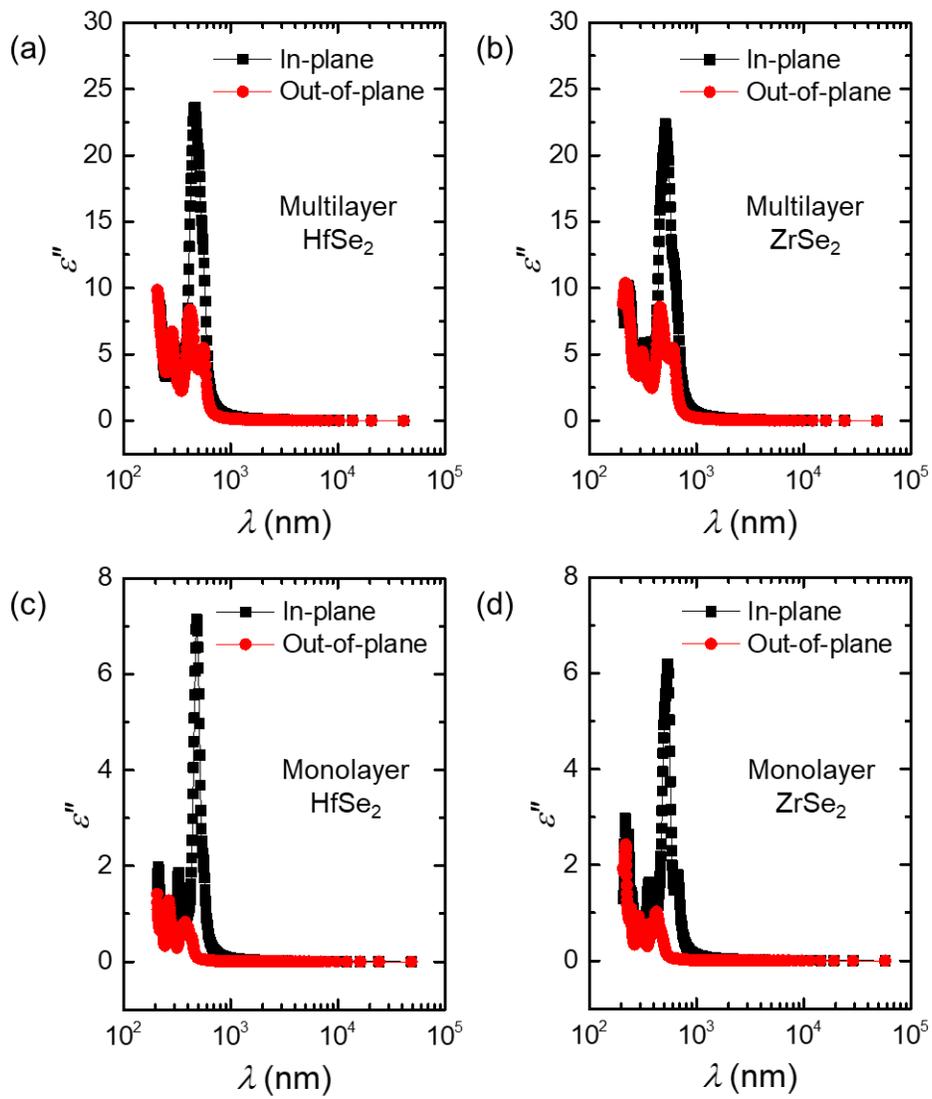

FIG. 4. Imaginary parts of dielectric constants of (a) multilayer $HfSe_2$, (b) multilayer $ZrSe_2$, (c) monolayer $HfSe_2$, and (d) monolayer $ZrSe_2$.

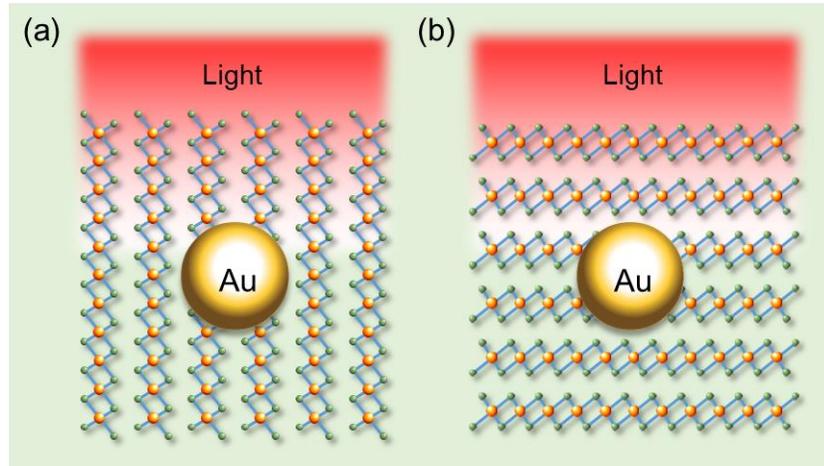

FIG. 5. Schematics of LSPR excited by an Au NP in the multilayer HfSe$_2$ or ZrSe$_2$ when the light illuminates along (a) in-plane direction and (b) out-of-plane direction.

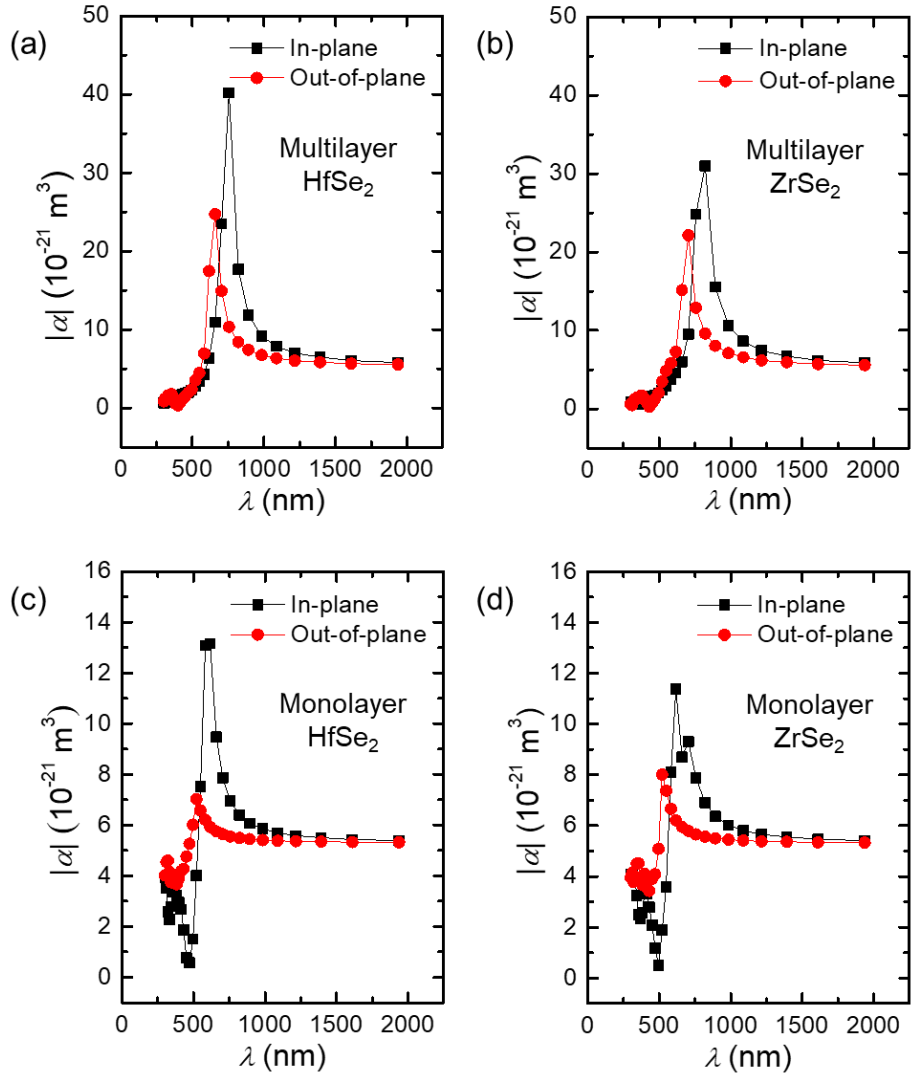

FIG. 6. Polarizabilities of an Au NP ($a = 75$ nm) as a function of wavelength on (a) multilayer HfSe$_2$, (b) multilayer ZrSe$_2$, (c) monolayer HfSe$_2$, and (d) monolayer ZrSe$_2$.

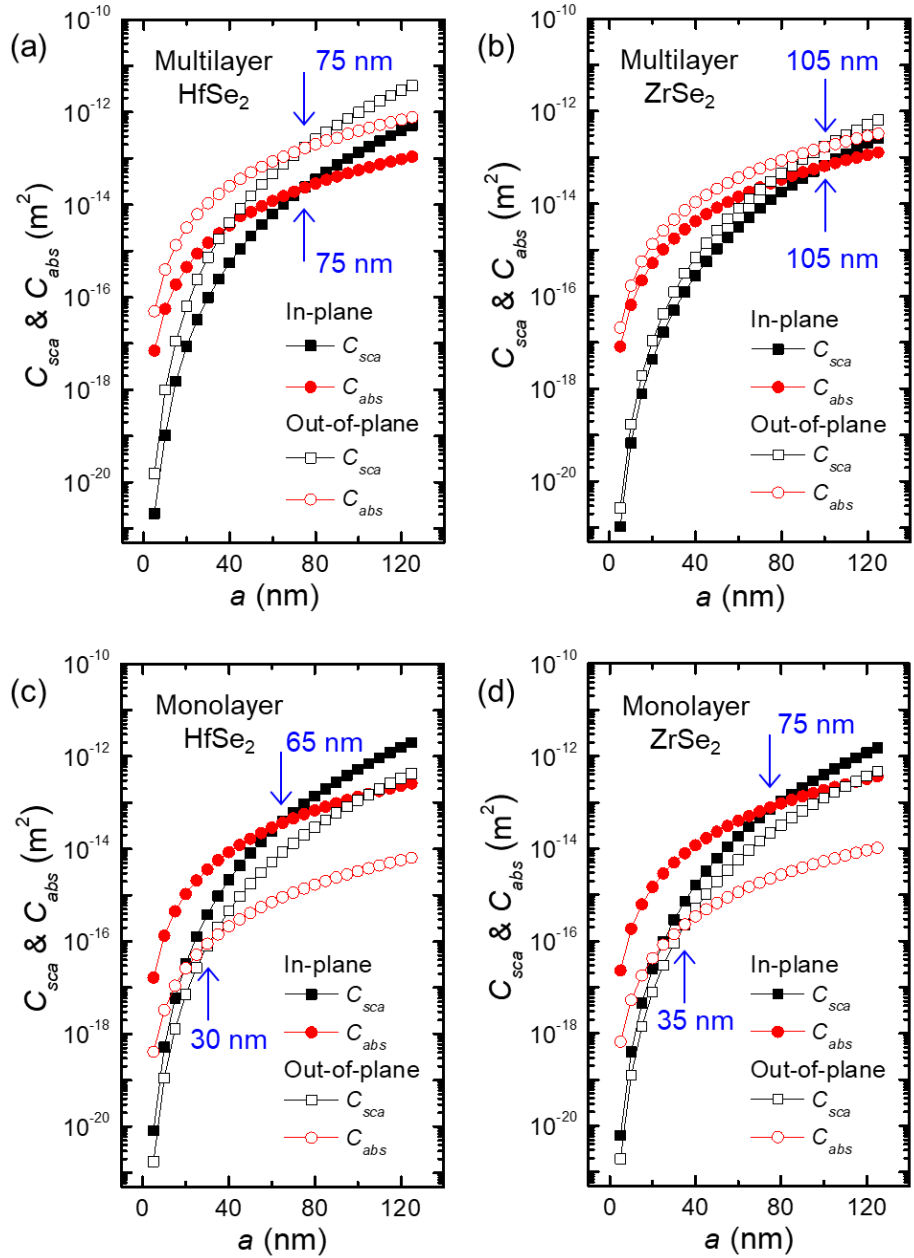

FIG. 7. Scattering and absorption cross-sections of an Au NP as a function of the NP radius in the environment of (a) multilayer $HfSe_2$, (b) multilayer $ZrSe_2$, (c) monolayer $HfSe_2$, and (d) monolayer $ZrSe_2$. The photon energy is 2 eV, and the blue arrow indicates the threshold radius.

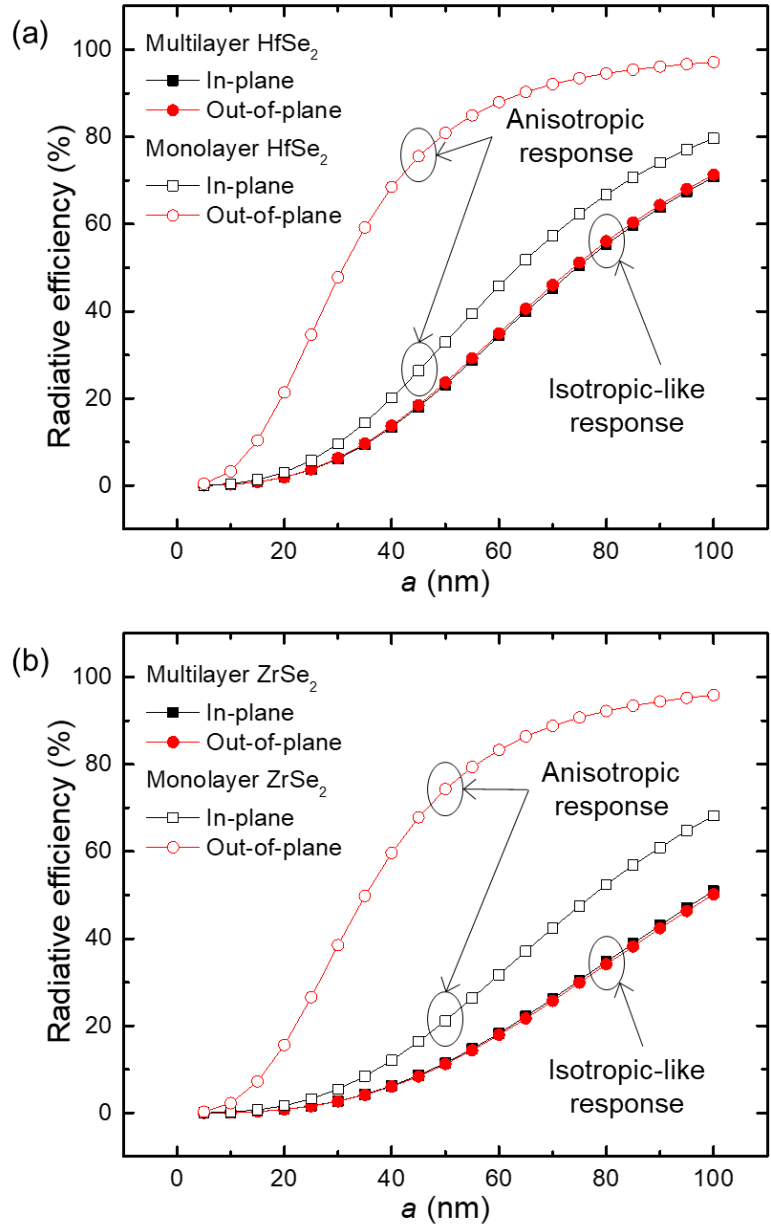

FIG. 8. Radiative efficiencies of an Au NP as a function of the NP radius on (a) HfSe$_2$ and (b) ZrSe$_2$. The photon energy is 2 eV.